\documentclass[
    ,final            
  ]
  {aipproc}

\layoutstyle{6x9}

\def\be{\begin{equation}}
\def\ee{\end{equation}}
\def\bea{\begin{eqnarray}}
\def\eea{\end{eqnarray}}
\def\bear{\begin{array}}
\def\ear{\end{array}}
\def\bfig{\begin{figure}}
\def\efig{\end{figure}}
\def\bcen{\begin{center}}
\def\ecen{\end{center}}
\def\bi{\begin{itemize}}
\def\ei{\end{itemize}}

\def\raw{\rightarrow}

\usepackage{amssymb}

\begin{document}

\title{Neutrino-induced coherent pion production}


\author{L. Alvarez-Ruso}{
  address={Departamento de F\'{\i}sica Te\'orica and IFIC, Universidad de Valencia - CSIC, Valencia, Spain}
}

\author{L. S. Geng}{
  address={Departamento de F\'{\i}sica Te\'orica and IFIC, Universidad de Valencia - CSIC, Valencia, Spain}
}

\author{S. Hirenzaki}{
  address={Department of Physics, Nara Women's University, Nara, Japan}
}

\author{M. J. Vicente Vacas}{
  address={Departamento de F\'{\i}sica Te\'orica and IFIC, Universidad de Valencia - CSIC, Valencia, Spain}
}

\author{T.~Leitner}{
  address={Institut f\"ur Theoretische Physik, Universit\"at Giessen, Germany}
}

\author{U. Mosel}{
  address={Institut f\"ur Theoretische Physik, Universit\"at Giessen, Germany}
}

\begin{abstract}
We have investigated the neutrino induced coherent pion production reaction
at the energies of interest for recent experiments like K2K and MiniBooNE. The model  
includes pion, nucleon and the $\Delta(1232)$ resonance. 
Medium effects in the production mechanism and the distortion of the pion wave
 function are taken into account. We find a strong reduction of the cross
section due to these effects and also substantial modifications in the
energy distributions of the final pion. The sensitivity of the results on the axial N-$\Delta$ 
coupling $C_5^A(0)$ and the coherent fraction in neutral-current $\pi^0$ production are discussed. 
\end{abstract}

\maketitle


The coherent production of pions in charged current (CC)  and neutral current (NC) processes 
is a subject of research in current and future experiments. The K2K 
collaboration has not found any evidence of $\nu_\mu + ^{12}C \raw \mu^- + \pi^+ + ^{12}C$, obtaining an 
upper limit for the coherent  fraction over the total CC interaction~\cite{Hasegawa:2005td} 
well below the estimates based on the Rein and Sehgal model~\cite{Rein:1982pf}. On the other side, preliminary 
MiniBooNE results indicate that part of the NC $\pi^0$ production comes from the 
coherent reaction  $\nu + ^{12}C \raw \nu + \pi^0 + ^{12}C$~\cite{JLink}. In future, the SciBooNE
detector~\cite{Mahn:2006ac} should be able to identify $\pi^0$'s emitted in the forward direction, 
where most of the coherent events are concentrated, while 
MINER$\nu$A~\cite{Drakoulakos:2004gn}
will collect data with high statistics, allowing for a clear separation between coherent and incoherent
processes and the comparison between neutrino and antineutrino cross sections.
 
Since the pioneering work of Ref.~\cite{Rein:1982pf} some other studies focused  on the 
energy region $\sim 1$~GeV, where the modification of
the $\Delta(1232)$ spectral function inside the nuclear medium is 
relevant~\cite{Kim:1996az,Kelkar:1996iv,Singh:2006bm}. Pion distortion is taken into account in 
Refs.~\cite{Rein:1982pf, Paschos:2005km} 
by factorizing the pion-nucleus elastic cross section (c.s.). 
In a more general fashion, it can be incorporated in the amplitude by means of the
distorted wave Born approximation, using a pion wave function obtained in the eikonal 
limit~\cite{Singh:2006bm} or by solving the Klein-Gordon equation with a realistic optical 
potential~\cite{Kelkar:1996iv}. 

We have performed a theoretical study of neutrino induced coherent 
pion~\cite{AlvarezRuso:2007tt,AlvarezRuso:2007it} production extending and improving
the calculations of Refs.~\cite{Kelkar:1996iv, Singh:2006bm}. 
The model is built in terms of the relevant hadronic degrees of freedom: 
pion, nucleon and $\Delta$ resonance.  Besides the dominant direct $\Delta$ excitation, 
it includes the crossed $\Delta$ and nucleon-pole terms~\cite{AlvarezRuso:2007it} 
(see the left panel of Fig.~\ref{fig12}). 
There are other contributions allowed by chiral symmetry~\cite{Hernandez:2007qq} but they cancel for 
isospin symmetric nuclei, so we neglect them.  

The relativistic amplitude is proportional to the 
product of the standard leptonic current and the nuclear current, obtained as the coherent sum over all 
nucleons. Detailed expressions of the different contributions to the nuclear current can be found in 
Ref.~\cite{AlvarezRuso:2007it}. The single-nucleon
contributions to the current are parametrized in terms of vector and axial form factors (FF).
The vector FF are related to the electromagnetic ones and can be extracted from electron scattering data. 
The axial FF are usually constrained by means of PCAC.    
For the $N-\Delta$ transition, this constraint is insufficient and the Adler model: 
$C_4^A = - C_5^A /4 $, $ C_3^A=0$ is adopted. For $C_5^A$ we consider two different parametrizations:
\be
\label{setI}
C_{5(I)}^A = {{C_5^A(0)\left[ 1+{{1.21\, q^2}/\left(2\,\,\mathrm{GeV}^2-q^2\right)} \right] }
{\left( 1- {{q^2}/{M_{A\Delta}^2}}\right)^{-2}}} \,,
\ee 
with $C_5^A(0) = 1.2$, in agreement with the off-diagonal Goldberger-Treiman (GT) relation, and
$M_{A\Delta}=1.28$~GeV, as extracted from BNL data, and
\be
\label{setII}
C_{5(II)}^A = C_5^A(0) \left(1 - q^2/3 M_{A\Delta}^2\right)^{-1} \left( 1- {{q^2}
/{M_{A\Delta}^2}}\right)^{-2} ,
\ee
with $C_5^A(0) = 0.867$ and $M_{A\Delta}=0.985$~GeV as fitted in 
Ref.~\cite{Hernandez:2007qq} to the ANL data with an invariant mass constraint of $W < 1.4$~GeV. 
Notice that most quark model calculations also obtain  $C_5^A(0)$ values that are smaller than 
the GT one (see Ref~\cite{BarquillaCano:2007yk} for a compilation). An  exception is the chiral 
quark model of Ref.~\cite{Golli:2002wy} where a fluctuating $\sigma$ field is taken into account. 

\bfig[hb!]
\begin{minipage}{.42\linewidth}
\includegraphics[width=0.99\textwidth]{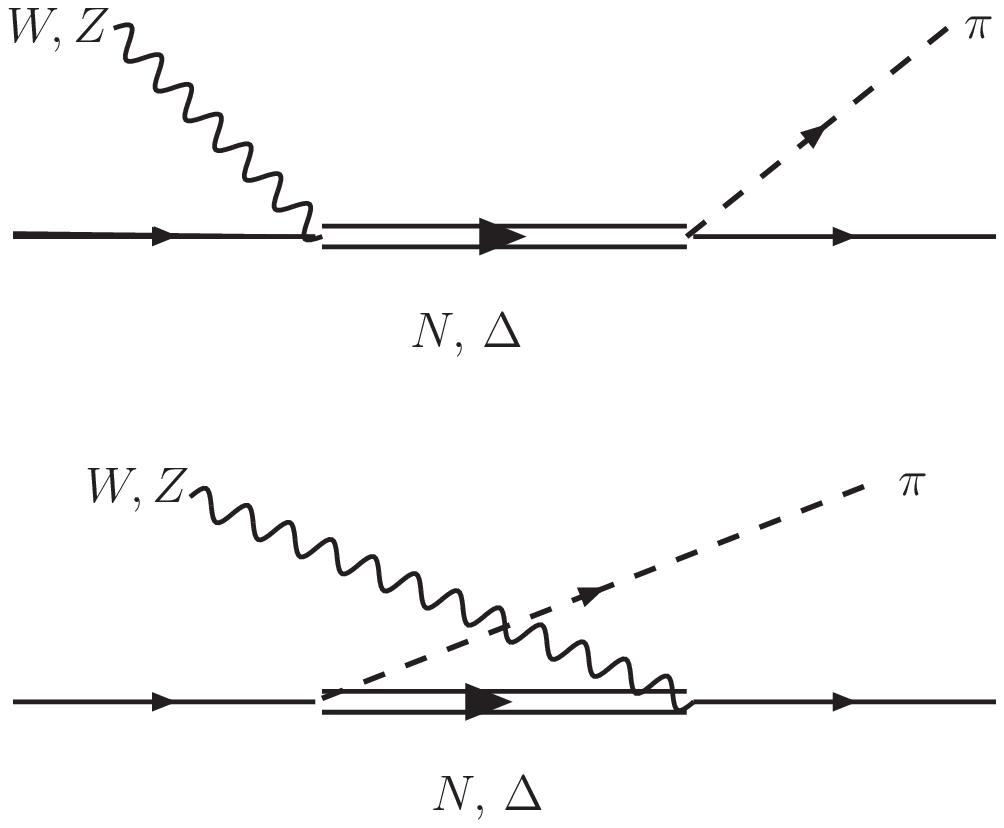}
\end{minipage}
\hfill
\begin{minipage}{.58\linewidth}
\includegraphics[width=0.99\textwidth]{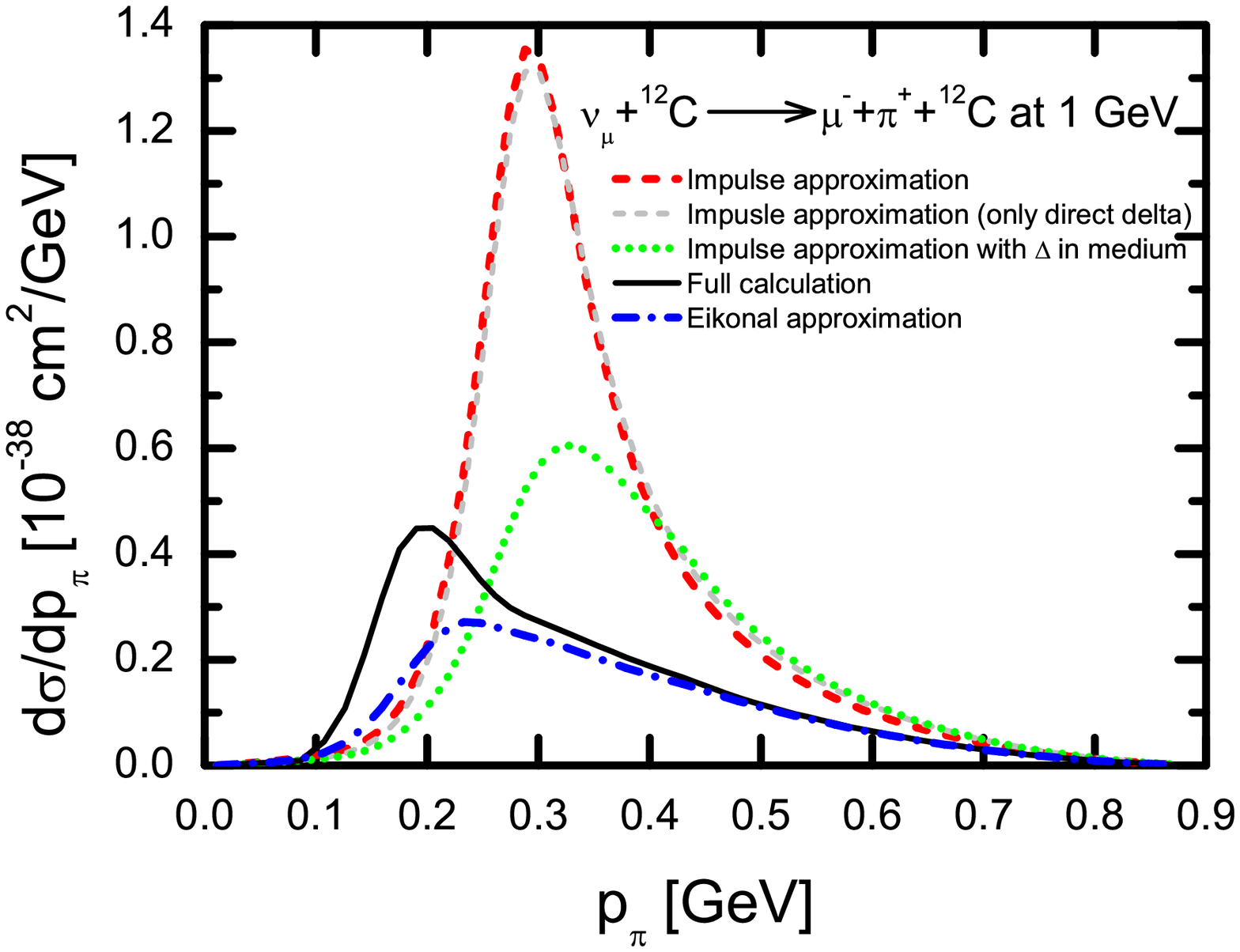}
\caption{{\it Left panel}: Elementary reaction mechanisms for coherent pion production on isospin-symmetric nuclei.
{\it Right panel}: Pion momentum distribution.}
\end{minipage} 
\label{fig12}
\efig

The pion momentum distribution for CC coherent $\pi^+$ production on $^{12}$C at $E_\nu = 1$~GeV is presented 
in Fig.~\ref{fig12}. The comparison of the two dashed lines shows that the
direct $\Delta$ excitation term accounts for most of the c.s. In fact, 
the crossed $\Delta$ amplitude is very small, and there is a cancellation between the direct and 
crossed pole-nucleon ones. 

The strong modification of the $\Delta$ properties inside the nuclear medium 
are taken into account  by adding a density dependent selfenergy.  This  reduces  the c.s. by around 35~\%.
A realistic quantum treatment of pion distortion can be achieved by solving the 
Klein-Gordon equation with a microscopic optical potential 
$\hat{V}_{opt}$~\cite{GarciaRecio:1989xa,Nieves:1991ye}
based on the $\Delta$-hole model.  Pion distortion further  decreases
the c.s. and moves the peak to lower energies. This reflects the presence of a strongly 
absorptive part in $\hat{V}_{opt}$ around the $\Delta$ peak. The eikonal 
approximation clearly fails at $p_\pi < 400$~MeV$/c$. 

\begin{table}
\caption{Cross sections for weak coherent pion production in units of $10^{-40}\textrm{cm}^2$,
averaged over the Aachen-Padova~\cite{Faissner:1977ku}, K2K~\cite{Ahn:2006zz}
and MiniBooNE~\cite{MiniBooNE} spectra.
$\sigma_\mathrm{I(II)}$ correspond to the form factors I and II.}
\label{table1}
\begin{tabular}{lccccc}
\hline 
Reaction & Experiment & $\sigma_\mathrm{I}$ & $\sigma_\mathrm{II}$  & $\sigma$ Experimental
\\
\hline
NC $\nu +\, ^{27}$Al      & Aachen-Padova   &  19.9 & 10.1 &  29$\pm$ 10 \cite{Faissner:1983ng}\\
NC $\bar{\nu}+\, ^{27}$Al &  Aachen-Padova  & 19.7 & 9.8 & 25$\pm$ 7 \cite{Faissner:1983ng}\\
CC $\nu+\, ^{12}$C        & K2K & 10.8 & 5.7 & $< 7.7$ \cite{Hasegawa:2005td}\tablenote{Obtained
using the ratio between coherent and $\sigma^{CC}$, the total CC cross section, and
 the value for $\sigma^{CC}$ of the K2K MC simulation.}\\
NC $\nu+\, ^{12}$C        & MiniBooNE  &  5.0   
& 2.6 & - \\
NC $\bar{\nu}+\, ^{12}$C & MiniBooNE  &  4.6 & 2.2  &-& \\
\hline
\end{tabular}
\end{table}

The c.s. averaged over the fluxes of Aachen-Padova, K2K and MiniBooNE experiments are given in 
Table~\ref{table1}. In the case of K2K the experimental threshold of 
$p_\mu >  450$~MeV/c is taken into account. 
The results obtained with set~II are about a factor two smaller than those 
obtained with set~I. This feature can be understood from the fact that in the forward direction 
($q^2 = 0$), where most of the strength of this reaction is
concentrated, the only form factor that contributes is $ C_5^A$~\cite{AlvarezRuso:1998hi}. 
Therefore, one can infer that 
$
\sigma (\mathrm{I})/\sigma (\mathrm{II}) \sim 
\left[ C_{5(\mathrm{I})}^A(0)/C_{5(\mathrm{II})}^A(0)\right]^2 \approx 1.9 
$. 
For Aachen-Padova, $\sigma_\mathrm{I}$ is below the central experimental values but 
within the large error bars, while with set II the experiment is clearly underestimated. 
On the contrary, for K2K only $\sigma_\mathrm{II}$ is below the experimental upper bound although 
one should bear in mind that nuclear effects may affect the experimental separation of  
coherent events from incoherent ones. The situation is illustrated on the left panel of Fig.~\ref{fig6} 
where we plot the muon angular distributions averaged over the K2K flux 
for coherent $\pi^+$ production, together with the main contributions to the total
inclusive CC cross section: quasielastic scattering (QE) and incoherent $\Delta$
 excitation.
The calculation of the $\Delta$ part is performed with set I. For the QE
 process, we have adopted the model of Ref.~\cite{Singh:1992dc}. 
Nuclear effects include Fermi motion, Pauli blocking and
the renormalization of the weak transition, treated as an
RPA resummation of particle-hole and $\Delta$-hole states. These nuclear correlations cause a 
considerable reduction of strength at low $q^2$ (forward angles), while they are negligible for
$\cos\theta_\mu < 0.8$. Therefore, if a model that lacks these correlations is used to 
extrapolate the data from the region of $\cos \theta_\mu \lesssim 0.8$ to forward angles, 
one might overestimate the QE part, causing an
underestimation of other mechanisms, like the coherent pion production.
\bfig[ht!]
\includegraphics[width=0.49\textwidth]{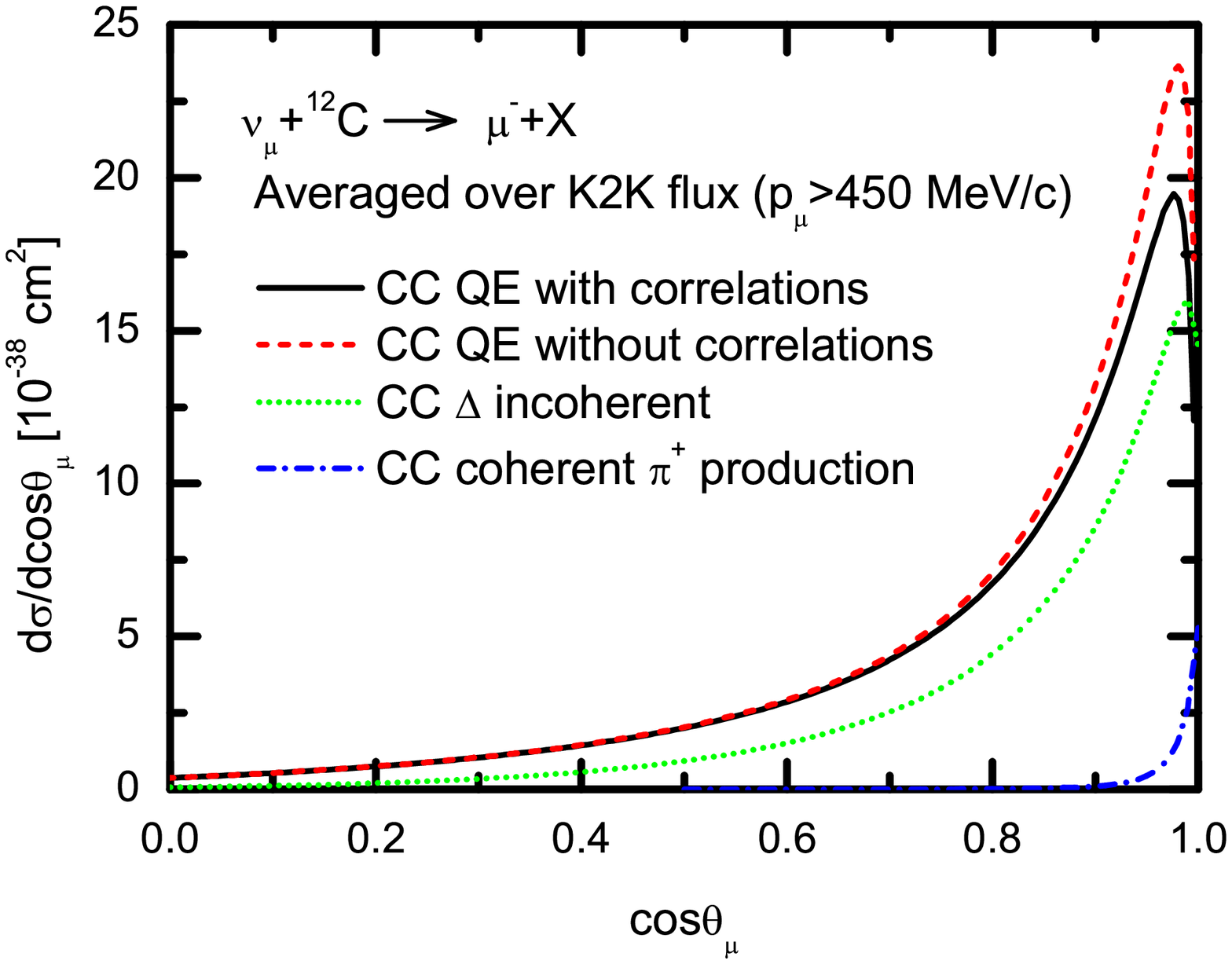}
\includegraphics[width=0.5\textwidth]{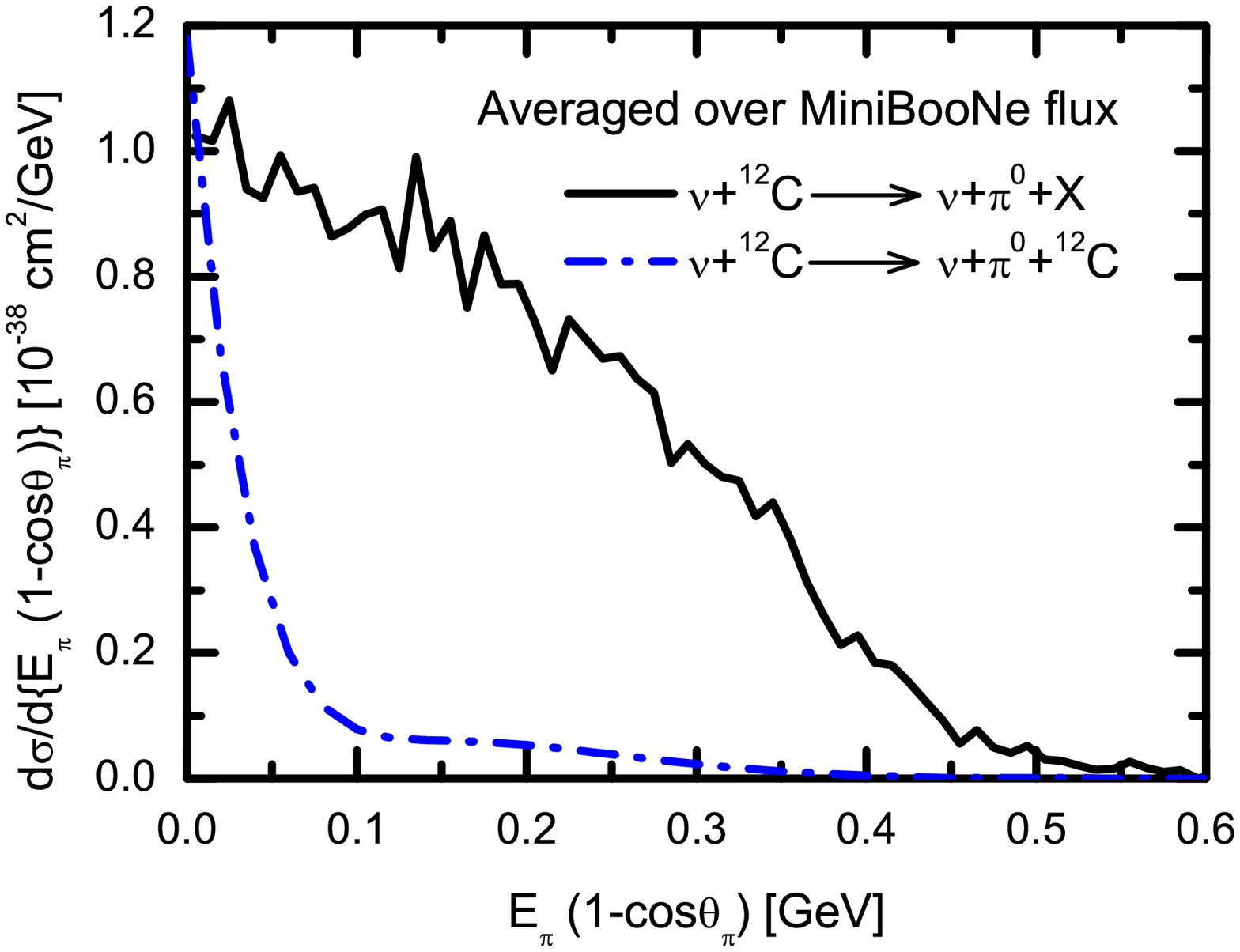}
\caption{{\it Left panel}: Different contributions to muon angular distribution for CC processes at K2K. 
{\it Right panel}: Coherent and incoherent contributions to total NC $\pi^0$ production at MiniBooNE.}
\label{fig6}
\efig
 
At MiniBooNE, the measured distribution in $E_\pi (1 - \cos\theta_\pi)$ is used to extract 
the coherent fraction from the total  NC $\pi^0$ production on $^{12}$C. For this reason we 
have performed new calculations of this observable for both coherent and incoherent $\pi^0$ 
production. The results are given on the right panel of Fig.~\ref{fig6}. 
In order to describe incoherent pion production, the interactions of the 
final particles inside the nucleus have to be properly taken into account, including quasielastic 
scattering, charge exchange and absorption (for pions). This is achieved with the semiclassical GiBUU 
transport model. The details of this model and an extensive set of results for $\nu$A scattering 
can be found in Refs.~\cite{Leitner:2006ww} and were presented by U. Mosel at this 
conference~\cite{Leitner:2007px}. The coherent fraction at MiniBooNE in our model is found to be 
\be
\frac{\sigma (\mathrm{coh.} )} {\sigma (\mathrm{coh.}) + \sigma(\mathrm{incoh.})} = 0.14 \,,
\ee
which is slightly below the preliminary value obtained by  MiniBooNE~\cite{JLink}.

\end{document}